\documentclass[10pt]{amsart}
\usepackage{amsfonts,amssymb}
\RequirePackage{ifpdf}
\usepackage{amsmath}
\usepackage{color}
\usepackage{pstcol,pst-node}
\usepackage{graphics}
\usepackage{epic}
\usepackage{curves}

\def\appendix#1{
\addtocounter{section}{1} \setcounter{equation}{0}
\renewcommand{\thesection}{\Alph{section}}
\section*{Appendix \thesection\protect\indent\quad
#1}
}
\renewcommand{\theequation}{\thesection.\arabic{equation}}

\catcode`\@=11
\def\marginnote#1{}

\newcount\hour
\newcount\minute
\newtoks\amorpm
\hour=\time\divide\hour by60 \minute=\time{\multiply\hour by60
\global\advance\minute by-\hour}
\edef\standardtime{{\ifnum\hour<12 \global\amorpm={am}%
        \else\global\amorpm={pm}\advance\hour by-12 \fi
        \ifnum\hour=0 \hour=12 \fi
        \number\hour:\ifnum\minute<10 0\fi\number\minute\the\amorpm}}
\edef\militarytime{\number\hour:\ifnum\minute<100\fi\number\minute}

\def\draftlabel#1{{\@bsphack\if@filesw {\let\thepage\relax
      \xdef\@gtempa{\write\@auxout{\string
          \newlabel{#1}{{\@currentlabel}{\thepage}}}}}\@gtempa \if@nobreak
    \ifvmode\nobreak\fi\fi\fi\@esphack} \gdef\@eqnlabel{#1}}
    \def\@eqnlabel{}
\def\@vacuum{}
\def\draftmarginnote#1{\marginpar{\raggedright\scriptsise\tt#1}}

\def\draft{
  \oddsidemargin -.5truein
  \def\@oddfoot{\footnotesise \sl preliminary draft \hfil
    \rm\thepage\hfil\sl\today\quad\militarytime}
  \let\@evenfoot\@oddfoot \overfullrule 3pt
    \let\label=\draftlabel
    \let\marginnote=\draftmarginnote
  \def\@eqnnum{(\theequation)\rlap{\kern\marginparsep\tt\@eqnlabel}%
    \global\let\@eqnlabel\@vacuum}

  }

\newcommand{\beq}{\begin{equation}}
\newcommand{\eeq}{\end{equation}}

\def\be{\begin{equation}}
\def\ee{\end{equation}}
\def\bea{\begin{eqnarray}}
\def\eea{\end{eqnarray}}
\def\<{\langle}
\def\>{\rangle}

\def\nn{\nonumber}

\def\one#1{#1^{\raise5pt\hbox{$\scriptstyle\!\!\!\!1$}}\,{}}
\def\two#1{#1^{\raise5pt\hbox{$\scriptstyle\!\!\!\!2$}}\,{}}
\def\onetwo#1{#1^{\raise5pt\hbox{$\scriptstyle\!\!\!\!\!{12}$}}\,{}}

\newtheorem{theorem}{Theorem}[section]

\theoremstyle{definition}

\newtheorem{remark}[theorem]{Remark}

\theoremstyle{remark}

\allowdisplaybreaks

\begin{document}
\title[{\it Confluence on the Painlev\'e monodromy manifolds and quantisation}]
{Confluence on the Painlev\'e monodromy manifolds, their Poisson structure and quantisation.}
\author{Marta Mazzocco, Vladimir Rubtsov}

\maketitle

\begin{abstract}
In this paper we obtain a system of flat coordinates on the monodromy manifold of each of the Painlev\'e equations. This allows us to quantise such manifolds. We produce a quantum confluence procedure between cubics in such a way that quantisation and confluence commute. We also investigate the underlying cluster algebra structure and the relation to the versal deformations of singularities of type $D_4,A_3,A_2$, and $A_1$. \end{abstract}


\section{Introduction}

Following the approach by Sakai \cite{sakai}, there are eight Painlev\'e equations corresponding to the eight extended Dynkin diagrams $\widetilde D_4,\widetilde D_5,\widetilde D_6,\widetilde D_7,\widetilde D_8,\widetilde E_6,\widetilde E_7,\widetilde E_8$, corresponding respectively to PVI, PV, three different cases of PIII, PIV, PII and PI. Their monodromy manifolds were studied by several authors, but were recently presented in  a unified way in \cite{SvdP}:
\begin{eqnarray}\nn
\widetilde D_4 & x_1 x_2 x_3 +x_1^2+x_2^2+x_3 ^2+\omega_1 x_1+\omega_2 x_2+\omega_3 x_3+\omega_4=0,\nn\\
&&\nn\\
\widetilde D_5 & x_1 x_2 x_3 +x_1^2+x_2^2+\omega_1 x_1+\omega_2 x_2+\omega_3 x_3+\omega_4=0,\nn\\
&&\nn\\
\widetilde D_6 & x_1 x_2 x_3 +x_1^2+x_2^2+\omega_1 x_1+\omega_2 x_2+\omega_1-1=0,\nn\\
&&\nn\\
\widetilde D_7 & x_1 x_2 x_3 +x_1^2+x_2^2+\omega_1 x_1=0,\nn\\
&&\nn\\
\widetilde D_8 & x_1 x_2 x_3 +x_1^2+x_2^2+1=0,\nn\\
&&\nn\\
\widetilde E_6 & x_1 x_2 x_3 +x_1^2+\omega_1 x_1+\omega_2 (x_2+ x_3)+1+\omega_4=0,\nn\\
&&\nn\\
\widetilde E_7^{\ast} & x_1 x_2 x_3 + x_1+ x_2+ x_3+\omega_4=0,\nn\\
&&\nn\\
\widetilde E_7^{\ast\ast} & x_1 x_2 x_3 + x_1+ \omega_2 x_2+ x_3-\omega_2+1=0,\nn\\
&&\nn\\
\widetilde E_8 & x_1 x_2 x_3+x_1+ x_2+1=0,\nn
\end{eqnarray}
where $\omega_1,\dots,\omega_4$ are some constants related to the parameters appearing in the Painlev\'e equations as described in Section \ref{se:unitary} here below and the two cubics $\widetilde E_7^\ast$ and $\widetilde E_7^{\ast\ast}$ correspond to the two different isomonodromy problems for PII found by Flaschka--Newell  \cite{FN} and Jimbo--Miwa  \cite{MJ1} respectively.

By looking at the above list of cubics it is immediately evident that one can follow the famous Painlev\'e confluence scheme (indeed the generalised one appearing in Sakai's paper \cite{sakai}) on the cubics by simple operations. For example, we can produce the PV  $\widetilde D_5$ cubic by considering the PVI  $\widetilde D_4$ one and rescaling $x_1\to\frac{x_1}{\epsilon }$, $x_2\to\frac{x_2}{\epsilon}$, $\omega_1\to\frac{\omega_1}{\epsilon}$, $\omega_2\to\frac{\omega_2}{\epsilon}$, $\omega_3\to\frac{\omega_3}{\epsilon^2}$ and $\omega_4\to\frac{\omega_4}{\epsilon^2}$ and then keeping the dominant term as $\epsilon\to 0$. This simple idea allows to us to extend the parameterisation of the PVI cubic in terms of shear coordinates obtained in \cite{ChM} to all other Painlev\'e equations. 

In particular, in this paper we study the natural Poisson bracket defined on these cubics, its relation with the log-canonical Poisson bracket, provide flat coordinates on each cubic and produce their quantisation. Interestingly we also produce a quantum confluence procedure in such a way that quantisation and confluence commute.

For the $\widetilde D_5,\widetilde D_6,\widetilde E_6,\widetilde E_7$ cubics we also associate a Riemann surface and its fat-graph to each cubic, so that our flat coordinates are indeed the Thurston shear coordinates on the fat--graph. Following the Fock--Goncharov philosophy, we also address the problem of whether there is some cluster algebra structure hidden in each cubic. We prove that indeed for 
$\widetilde D_4, \widetilde D_5, \widetilde D_6$ and $\widetilde E_6$ there is a {\it tagged cluster algebra}\/ structure \cite{ChS}.  In particular this implies that the procedure of analytic continuation of the solutions to the sixth Painlev\'e equation satisfies the Laurent phenomenon as explained in Section \ref{se:cluster}.

Last but not least, we interpret each Painlev\'e cubic as versal deformation of a Arnold singularity according to Sakai's table.

This paper is organised as follows: in Section \ref{se:unitary}, we recall the link between the parameters  $\omega_1,\dots,\omega_4$ and the Painlev\'e parameters $\alpha,\beta,\gamma$ and $\delta$. In Section \ref{se:volume-forms} we discuss the Cayley cubic and its relation to the log-canonical Poisson bracket and we interpret each Painlev\'e cubic as versal deformation of a Arnold singularity according to Sakai's table. In Section \ref{se:cluster} we explain the tagged cluster algebra structure appearing in the case of PVI, PV, PIII and PIV. In Section \ref{se:shear} we present the flat coordinates for each cubic. In Section \ref{se:q} we present the quantisation and the quantum confluence.

\vskip 2mm \noindent{\bf Acknowledgements.} The authors are grateful to
Yu. Berest, O. Chalyck, L. Chekhov, P. Clarkson,  F. Eshmatov, V. Sokolov and T. Sutherland  for helpful discussions. We are thankful to B.V. Dang for his help with SINGULAR package. This research was supported by the EPSRC Research Grant $EP/J007234/1$, by the Hausdorff Institute, by ANR "DIADEMS", by RFBR-12-01-00525-a,  MPIM (Bonn) and SISSA (Trieste).

\section{Unified approach to the monodromy manifolds}\label{se:unitary}

According to \cite{SvdP}, the monodromy manifolds $\mathcal M^{(d)}$ have all the form
\be\label{eq:mon-mf}
x_1 x_2 x_3 + \epsilon_1^{(d)} x_1^2+ \epsilon_2^{(d)} x_2^2+ \epsilon_3^{(d)} x_3^2 + \omega_1^{(d)} x_1  + \omega_2^{(d)} x_2 + \omega_3^{(d)} x_3+
\omega_4^{(d)}=0,
\ee
where $d$ is an index running on the list of the extended Dynkin diagrams $\widetilde D_4,\widetilde D_5,\widetilde D_6$, $\widetilde D_7,\widetilde D_8,\widetilde E_6,\widetilde E_7^\ast,\widetilde E_7^{\ast\ast},\widetilde E_8$  and the parameters $ \epsilon^{(d)}_{i},\, \omega^{(d)}_{i}$, $i=1,2,3$ are given by:
\bea\label{eq:epsilon}
&&
 \epsilon^{(d)}_{1} =\left\{\begin{array}{ll}
 1&\hbox{ for } d=\widetilde D_4,\widetilde D_5,\widetilde D_6,\widetilde D_7,\widetilde D_8,\widetilde E_6,\\
0&\hbox{ for } d=\widetilde E_7^\ast,\widetilde E_7^{\ast\ast},\widetilde E_8,\\
 \end{array}\right.\nn\\
 &&
 \epsilon^{(d)}_{2} =\left\{\begin{array}{ll}
 1&\hbox{ for } d=\widetilde D_4,\widetilde D_5,\widetilde D_6,\widetilde D_7,\widetilde D_8\\
0&\hbox{ for } d= \widetilde E_6,\widetilde E_7^\ast,\widetilde E_7^{\ast\ast},\widetilde E_8,
 \end{array}\right.\\
 &&
  \epsilon^{(d)}_{3} =\left\{\begin{array}{ll}
 1&\hbox{ for } d=\widetilde D_4,\\
0&\hbox{ for } d=\widetilde D_5,\widetilde D_6,\widetilde D_7,\widetilde D_8,\widetilde E_6,\widetilde E_7^\ast,\widetilde E_7^{\ast\ast},\widetilde E_8.
 \end{array}\right.\nn
 \eea
 while 
 \bea\label{eq:omega}
 &&
 \omega^{(d)}_{1} =
-G_1^{(d)}G_\infty^{(d)}-\epsilon^{(d)}_1 G_2^{(d)} G_3^{(d)} ,\quad \omega^{(d)}_{2} =
-G_2^{(d)}G_\infty^{(d)}-\epsilon^{(d)}_2 G_1^{(d)} G_3^{(d)},\nn\\
&&
 \omega^{(d)}_{3} =
-G_3^{(d)}G_\infty^{(d)}-\epsilon^{(d)}_3 G_1^{(d)} G_2^{(d)},\\
&&
 \omega^{(d)}_{4} =\epsilon^{(d)}_2\epsilon^{(d)}_3\left(G_1^{(d)}\right)^2+\epsilon^{(d)}_1\epsilon^{(d)}_3\left(G_2^{(d)}\right)^2
 +\epsilon^{(d)}_1\epsilon^{(d)}_2\left(G_3^{(d)}\right)^2+\left(G_\infty^{(d)}\right)^2+\nn\\
 &&\qquad\quad+G_1^{(d)} G_2^{(d)}
 G_3^{(d)} G_\infty^{(d)}-4\epsilon^{(d)}_1\epsilon^{(d)}_2\epsilon^{(d)}_3,\nn
 \eea
 where $G_1^{(d)}, G_2^{(d)}, G_3^{(d)}, G_\infty^{(d)}$ are some constants related to the parameters appearing in the Painlev\'e equations as follows:
 $$
 G_1^{(d)}=\left\{\begin{array}{lc}
 2\cos\pi\theta_0& d=\widetilde D_4,\widetilde D_5, \widetilde E_6\\ 
  e^{-\frac{i\pi(\theta_0+1)}{2}} &d=\widetilde E_7^\ast\\
    e^{-i\pi\theta_0} &d=\widetilde E_7^{\ast\ast}\\
    1&d=\widetilde D_7,\widetilde D_8,\widetilde E_8\\
   e^{\frac{i\pi(\theta_0+\theta_\infty)}{2}}+e^{\frac{-i\pi(\theta_0+\theta_\infty)}{2}}& d=\widetilde D_6,\\
 \end{array}\right.
 $$
 $$
 G_2^{(d)}=\left\{\begin{array}{lc}
 2\cos\pi\theta_1& d=\widetilde D_4,\widetilde D_5, \\ 
 2\cos\pi\theta_\infty& d=\widetilde E_6 \\
 e^{-\frac{i\pi(\theta_0+1)}{2}} &d=\widetilde E_7^\ast\\
  e^{i\pi\theta_0} &d=\widetilde E_7^{\ast\ast}\\
      1&d=\widetilde D_8,\widetilde E_8\\
  e^{\frac{i\pi(\theta_0-\theta_\infty)}{2}}+e^{\frac{i\pi(-\theta_0+\theta_\infty)}{2}}&d=\widetilde D_6\\
 \end{array}\right.
 $$
  $$
 G_3^{(d)}=\left\{\begin{array}{lc}
 2\cos\pi\theta_t& d=\widetilde D_4, \\ 
 1& d=\widetilde D_5,\widetilde D_7\\
 2\cos\pi\theta_\infty& d=\widetilde E_6 \\
  e^{-\frac{i\pi(\theta_0+1)}{2}} &d=\widetilde E_7^\ast\\
    e^{-i\pi\theta_0} &d=\widetilde E_7^{\ast\ast}\\
        0&d=\widetilde D_6,\widetilde D_8, \widetilde E_8\\
 \end{array}\right.
 $$
 $$
 G_\infty^{(d)}=\left\{\begin{array}{lc}
 2\cos\pi\theta_\infty& d=\widetilde D_4,\widetilde D_5,\widetilde E_6 \\ 
  e^{\frac{i\pi(\theta_0+1)}{2}} &d=\widetilde E_7^\ast\\
    e^{i\pi\theta_0} &d=\widetilde E_7^{\ast\ast}\\
        1&d=\widetilde D_8,\widetilde E_8\\
  e^{\frac{i\pi(\theta_0+\theta_\infty)}{2}}&d=\widetilde D_6\\
  0&d=\widetilde D_7\\
  \end{array}\right.
 $$
 where the parameters $\theta_0,\theta_1,\theta_t,\theta_\infty$ are related to the Painlev\'e equations parameters in the usual way \cite{MJ1}. 
  
 \begin{remark}
Observe that in the original article  \cite{SvdP} the $\widetilde E_7^\ast$ cubic corresponding to the Flaschka--Newell isomonodromic problem \cite{FN} has different signs. This can be obtained from our cubic (\ref{eq:mon-mf}) by a simple sign change: $x_2\to -x_2$
Analogously the  $\widetilde E_8$ and $\widetilde D_6$ cubics have different signs in \cite{SvdP}, which can both be obtained by the same sign change  $x_i\to-x_i$ for $i=1,2$. Finally to obtaine the $\widetilde D_8$ cubic as in \cite{SvdP} we just need to rescale $x_1\to i x_1$ and $x_3\to i x_3$. 
 \end{remark}
 
  \begin{remark}
 The cubic family of the monodromy manifolds $\mathcal M^{(d)}$  \ref{eq:mon-mf} type  appears in many different contexts. The $\widetilde D_4-$case was studied in 
Oblomkov' s work (see \cite{Obl}). W.Goldman and D. Toledo (\cite{GT})  had proved  that every cubic surface with $\epsilon^{d}_i = 1$  for all $i=1,2,3$ ( and at least one $\omega^{d}_i \neq 0$ for $i=1,2,3$ arises from a representation of the fundamental group of the 4-holed sphere in $SL(2,\mathbb C)$. They also have shown  that if  all $\omega^{d}_i = 0$ for $i=1,2,3$ and $\omega_4 \neq 0$ then the 4-holed sphere  should be replaced by 1-hole torus.  In the Painlev\'e context  the family of surfaces were considered by S. Cantat et F. Loray (\cite{CantLor}) and  M. Inaba, K. Iwasaki and M.Saito in \cite{IIS}. The first author (together with L. Chekhov) has studied the shear coordinates on $D_4-$ type family in the paper \cite{ChM}. We want to mention also that  M. Gross, P. Hacking and S.Keel (see Example  5.12 of \cite{GrossHK}) claim that  the family
 \ref{eq:mon-mf} can be "uniformize" by some analogues of theta-functions related to toric mirror data on log-Calabi-Yau surfaces.

 \end{remark}

\section{A digression on volume forms, singularities and different Dynkin diagrams}\label{se:volume-forms}

We would like to address here some natural facts that arise when comparing the various descriptions of 
 family of affine cubics surfaces with 3 lines at infinity  (\ref{eq:mon-mf}).

First of all,  the projective completion of the family of cubics \ref{eq:mon-mf} with $\epsilon^{(d)}_i \neq 0$ for all $ i = 1,2,3$ has singular points
only in the finite part of the surface and  if any of  $\epsilon^{(d)}_i, i = 1,2,3$ vanish, then $\mathcal M^{(d)}$ is singular at infinity with singular points in 
homogeneus coordinates $X_i =1$ and $X_j = 0, j\neq i$ (\cite{Obl}). Here $x_i=\frac{X_i}{X_0}$.

One can consider this family of cubics as a variety ${\mathcal S} =\{(\bar x,\bar\omega)\in \mathbb C^3 \times\Omega): S(\bar x,\bar \omega)=0\}$ where $\bar x=(x_1, x_2, x_3),\quad \bar\omega = (\omega_1,\omega_2, \omega_3, \omega_4)$ and the "$\bar x-$forgetful" projection $\pi :{\mathcal S}\to \Omega :
\pi(\bar x, \bar \omega) = \bar\omega.$ This projection defines a family of affine cubics with generically non--singular fibres $\pi^{-1}( \bar\omega)$ (we will discuss the nature of these singularities in Subsection \ref{se:sing}).

The cubic surface $S_{\bar \omega}$ has a volume form $\vartheta_{\bar \omega}$ given by the 
Poincar\'e residue formulae:
\be\label{eq:volform}
\vartheta_{\bar \omega}=\frac{dx_1\wedge dx_2}{(\partial S_{\bar \omega})/(\partial x_3)}=
\frac{dx_2\wedge dx_3}{(\partial S_{\bar \omega})/(\partial x_1)}=\frac{dx_3\wedge dx_1}{(\partial S_{\bar \omega})/(\partial x_2)}.
\ee

The volume form is a holomorphic 2-form on the non-singular part of $S_{\bar \omega}$ and it
has singularities along the singular locus. This form defines  the Poisson brackets on the surface
in the usual way as

\be\label{eq:poiss}
\{x_1, x_2\}_ {\bar \omega}=\frac{\partial S_{\bar \omega}}{\partial x_3}
\ee
and the other brackets are defined by circular transposition of $x_1, x_2, x_3.$ 
It is a straightforward computation to show that for $(i,j,k)=(1,2,3)$:

\be\label{eq:poiss}
\{x_i, x_j\}_ {\bar \omega}=\frac{\partial S_{\bar \omega}}{\partial x_k} = x_i x_j +2\epsilon_{i}^d x_k +\omega_{i}^d
\ee
and the volume form (\ref{eq:volform}) reads as
\be\label{eq:volform1}
\vartheta_{\bar \omega}=\frac{dx_i\wedge dx_j}{(\partial S_{\bar \omega})/(\partial x_k)}=
\frac{dx_i\wedge dx_j}{(x_i x_j +2\epsilon_{i}^d x_k +\omega_{i}^d)}.
\ee

In a special case of PVI, i.e. the $\widetilde D_4$ cubic with parameters $\omega_i=0$ for $i=1,2,3$ and $\omega_4=-4$, there is an isomorphism $\pi : \mathbb C^*\times \mathbb C^*/\eta \to S_{\bar \omega}:$ \cite{CantLor} 
\be\label{iso}
\pi (u,v) \to (x_1,x_2,x_3)=(-u-1/u,-v-1/v,-uv-1/uv),
\ee 
where $\eta$ is the involution of $\mathbb C^*\times \mathbb C^*$ given by
$u\to 1/u,\quad v\to 1/v.$ The log-canonical 2-form  $\bar\vartheta = \frac{du\wedge dv}{uv}$ defines a symplectic structure on $\mathbb C^*\times \mathbb C^*$ which is invariant with respect the involution $\eta$ and therefore defines a symplectic structure on the non-singular part of the cubic surface $S_{\bar\omega}$ for  $\omega_i=0$ for $i=1,2,3$ and $\omega_4=-4$. 

The relation between the log-canonical 2-form $\bar\vartheta = \frac{du\wedge dv}{uv}$ and the Poisson brackets on the surface $S_{\bar \omega}$ can be extended to all values of the parameters $\bar\omega$ and for all the 
Painlev\'e cubics as we shall show in this paper. In fact the flat coordinates that we will introduce in Section \ref{se:shear} are such that their exponentials satisfy the log-canonical Poisson bracket. Before doing so, we clarify  the relation between the Painlev\'e cubics and singularity theory.

\subsection{Singularity theory approach to the Painlev\'e cubics}\label{se:sing}

As mentioned above,  for special values of $\omega_1^{(d)},\dots,\omega_4^{(d)}$ the fibre  may have a singularity. Such singularities were classified in \cite{IIS}  for PVI and in \cite{SvdP} for all other Painlev\'e equations. These results can be summarised in the following table:

\begin{table}[h]
\begin{center} 
\begin{tabular}{|c|c||c|c|} \hline 
Dynkin & Painlev\'e equations & Surface singularity type \\ \hline 
 $\widetilde D_4$ & $P_{VI}$ &  $D_4$ \\ \hline 
$\widetilde D_5$ & $P_{V}$ & $A_3$ \\ \hline 
$\widetilde D_6$  &  deg $P_V$=$P_{III}(\widetilde D_6)$ & $A_1$ \\ \hline 
$\widetilde D_6$ & $P_{III}(\widetilde D_6)$ & $A_1$\\ \hline  
$\widetilde D_7$  & $P_{III}(\widetilde D_7)$ & non-singular \\ \hline 
$\widetilde D_8$  & $P_{III}(\widetilde D_8)$ & non-singular \\ \hline  
$\widetilde E_6$ &$ P_{IV}$     &  $A_2$ \\ \hline
$\widetilde E_7^{*}$ & $P_{II}(FN)$ & $A_1$ \\ \hline  
$\widetilde E_7^{**}$ & $P_{II}(MJ)$ & $A_1$ \\ \hline 
$\widetilde E_8$ & $P_{I}$  & non-singular \\ \hline 
\end{tabular}
\vspace{0.2cm}
\end{center}
\caption{}
\label{tab:sing}
\end{table}

The meaning of the table is the following: for each Painlev\'e equation of type specified by the first column in the table,  there is at least one singular fibre with singularity of the type given in the second column of the table, and at least one singular fibre with singularity of type specified by any Dynkin sub-diagram of the Dynkin diagram given in the second column of the table. 

For example PIV is the equation corresponding to $\widetilde {E_6}$ and it has a two singular fibres with singularity of type $A_2$ and at three singular fibres with singularity of type $A_1$.

The scope of this section is to show that the non singular fibres of each family of affine cubics are locally diffeomorphic to the versal unfolding \cite{Arnold} of the singularity of the type given in the second column of the table.

\subsubsection{$\widetilde D_4$}
This case corresponds to the sixth Painlev\'e equation. The cubic in this case is (we drop the indices ${(\widetilde D_4)}$ for convenience):
\be\label{eq:mon-mf-PVI}
x_1 x_2 x_3 + x_1^2+  x_2^2+  x_3^2 + \omega_1 x_1  + \omega_2  x_2 + \omega_3 x_3+
\omega_4 =0.
\ee
To show that this is diffeomorphic to the versal unfolding of $D_4$ we need to map this cubic to Arnol'd form. To this aim we first shift all variables by $x_i\to x_i +2$, $i=1,2,3$ to obtain
\be\label{eq:PVI-shifted}
x_1^2+x_2^2+x_3^2 + 2 x_1 x_2+ 2 x_2 x_3 + 2 x_1 x_3 + x_1 x_2 x_3 + \widetilde\omega_1 x_1  + \widetilde\omega_2  x_2 + \widetilde\omega_3 x_3+\widetilde\omega_4 =0,
\ee
where 
$$
 \widetilde\omega_i = \omega_i + 8, \quad\hbox{for } i=1,2,3,\qquad
  \widetilde\omega_4= \omega_4+2 (\omega_1+\omega_2+\omega_3)+20.
 $$
As a second step we use the following diffeomorphism around the origin:
$$
x\to x-\frac{1}{2} y, \quad y\to x+\frac{1}{2} x, \quad z\to z+\frac{y^2}{8}-2 x-\frac{x^2}{2} -\frac{\widetilde\omega_3}{2}
$$
so that the new cubic (up to a Morse singularity that we throw away and after a shift $x\to x-\frac{\omega_3}{4}$) becomes indeed the versal unfolding of a $D_4$ singularity in Arnol'd form:
$$
-2 x_1^3+ \frac{x_1 x_2^2}{2} +\widehat\omega_1 x_1+\widehat\omega_2 x_2+ \widehat\omega_3 x_1^2+\widehat\omega_4,
$$
where 
\bea
&&
\widehat\omega_1 = \omega_1+\omega_2 -8-4 \omega_3-\frac{\omega_3^2}{8}, \quad
\widehat\omega_2= \frac{\omega_2-\omega_1}{2},\nn\\
&&
\widehat\omega_3=8+\omega_3,\quad
\widehat\omega_4= \omega_4+2\omega_3-\frac{\omega_3(\omega_1+\omega_2-\omega_3)}{4}+4.\nn
 \eea
The above formulae show that the versal unfolding parameters $\widehat\omega_1,\dots,\widehat\omega_4$ are independent as long as $\omega_1,\dots,\omega_4$ are.

\subsection{$\widetilde D_5$}
This case corresponds to the fifth Painlev\'e equation. The cubic in this case is (we drop the indices ${(\widetilde D_5)}$ for convenience):
\be\label{eq:mon-mf-PV}
x_1 x_2 x_3 + x_1^2+  x_2^2+ \omega_1 x_1  + \omega_2  x_2 + \omega_3 x_3+
\omega_4 =0,
\ee
where only three parameters are free:
$$
\omega_1=-G_1 G_\infty-G_2,\quad \omega_2=-G_2 G_\infty-G_1,\quad \omega_3=- G_\infty,\quad 
\omega_4=1+G_\infty^2+G_1 G_2 G_\infty.
$$
Again we want to show that this is diffeomorphic to the versal unfolding of $A_3$. To this aim we impose the following change of variables: 
\be\label{transf}
x_1\to x_1-x_3+\frac{G_\infty}{u(x_2)}, \quad x_2\to u(x_2),\quad
x_3\to 2\frac{x_3}{u(x_2)}+\frac{G_2 + G_1 G\infty}{u(x_2)}-\frac{2 G_\infty}{u(x_2)^2}, 
\ee
where $u(x_2)$ is a function to be determined. This maps the PV cubic to:
$$
x_1^2-x_3^2 + 1+G_1 G_2 G_\infty+ G_\infty^2+ \frac{G_\infty^2}{u^2}-\frac{G_\infty(G_2+G_1 G_\infty)}{u}-(G_1+G_2 G_\infty) u + u^2.
$$
It is easy to prove that any solution $u(x_2)$ of the equation
$$
 \frac{G_\infty^2}{u^2}-\frac{G_\infty(G_2+G_1 G_\infty)}{u}-(G_1+G_2 G_\infty) u + u^2= x_2^4 + (G_2+G_1 G_\infty)x_2 ^2+ (G_1+G_2 G_\infty)x_2
 $$
will define a diffeomorphism by (\ref{transf}) mapping (\ref{eq:mon-mf-PV}) to the versal unfolding of $A_3$.

\subsection{$\widetilde D_6$}
This case corresponds to the third Painlev\'e equation. The cubic in this case is (we drop the indices ${(\widetilde D_3)}$ for convenience):
\be\label{eq:mon-mf-PV}
x_1 x_2 x_3 + x_1^2+  x_2^2+ \omega_1 x_1  + \omega_2  x_2+
\omega_4 =0,
\ee
where only two parameters are free:
$$
\omega_1=-1- G_\infty^2,\quad \omega_2=-G_2 G_\infty,\quad 
\omega_4=G_\infty^2.
$$
The most singular fibre is given by $G_\infty=1$ and $G_2=2$ and has two singular points at $(1, 0, 2)$ and $(0, 1, 2)$ respectively. We can define two local diffeomorphisms, one around $(1, 0, 2)$, the other around $(0, 1, 2)$, which map our cubic to the versal unfolding of  a $A_1$ singularity. 

The first diffeomorphism is given by: 
$$
x_1\to \frac{1+G_\infty^2}{2}+x_1,\quad 
x_2\to -x_2+ x_3 ,\quad x_3\to \frac{2(G_2 G_\infty-2 x_3   )}{1+ G_\infty^2+2 x_1}
$$
The second diffeomorphism is:
$$
x_1\to -x_1+x_3,\quad 
x_2\to \frac{G_2G_\infty}{2}-x_2 ,\quad x_3\to \frac{2( 1+ G_\infty^2 -2 x_3 )}{G_2 G_\infty^2-2 x_2}.
$$

\subsubsection{$\widetilde E_6$} 
This case corresponds to the fourth Painlev\'e equation. The cubic in this case is (we drop the indices ${(\widetilde E_6)}$ for convenience):
\be\label{eq:mon-mf-PIV}
x_1 x_2 x_3 + x_1^2+ \omega_1 x_1  + \omega_2  x_2 + \omega_3 x_3+
\omega_4 =0,
\ee
where only two parameters are free:
$$
\omega_1=-G_1 G_\infty-G_\infty^2,\quad \omega_2=- G_\infty^2,\quad \omega_3=- G_\infty^2,\quad 
\omega_4=G_\infty^2+G_1 G_\infty^3.
$$
Again we want to show that this is diffeomorphic to the versal unfolding of $A_2$. To this aim we impose the following change of variables: 
\be\label{transf1}
x_1\to  x_1-x_3+\frac{G_\infty^2}{u},\quad 
x_2\to u,\quad x_3\to \frac{2 x_3}{u}+ \frac{G_\infty}{u}(G_1+G_\infty)-\frac{2 G_\infty^2}{u^2}
\ee
where $u$ is  function of $x_3$ satisfying the following 
$$
\frac{G_\infty^4}{u^2} -\frac{G_\infty^3(G_\infty+G_1)}{u}-G_\infty^2 u= x_2^3+ G_\infty x_2.
$$
It is easy to prove that this transformation is a local diffeomorphism mapping our cubic to 
$$
 x_1^2-x_3^2 +x_2^3+ G_\infty x_2+G_\infty+G_1 G_\infty^3,
$$
the versal unfolding of the $A_2$ singularity.

\subsubsection{$\widetilde E_7$}
This case corresponds to the second Painlev\'e equation. Since the treatment of the two cubics $\widetilde E_7^\ast$ and $\widetilde E_7^{\ast\ast}$ is completely equivalent, we choose to work with the former:
\be\label{eq:mon-mf-PII}
x_1 x_2 x_3 - x_1  - x_2 -  x_3+
\omega_4 =0,
\ee
where:
$$
\omega_4=G_\infty^2+G_\infty^{-2}
$$
The following change of variables: 
$$
x_1\to  x_1-x_3+\frac{1}{u}, \quad x_2\to u, \quad 
x_3\to \frac{x_1+ x_3+1}{u},
$$
where $u$ is a function of $x_2$ satisfying 
$$
-\frac{1}{u}-u= x_2^2,
$$
is a local diffeomorphism mapping our cubic to the versal unfolding of the $A_1$ singularity:
$$
x_1^2-x_3^2+x_2^2+\omega_4.
$$

\section{Painlev\'e VI: analytic continuation and cluster mutations}\label{se:cluster}

In \cite{DM,M1} it was proved that the procedure of analytic continuation of a local solution to the sixth Painlev\'e equation corresponds to the following action of the braid group on the monodromy manifold:
\be\label{eq:braid1}
\beta_1:\begin{array}{lcl}
x_1&\to&-x_1-x_2x_3-\omega_1,\\
x_2&\to&x_3,\\
x_3&\to&x_2,\\
\end{array}\ee
\be\label{eq:braid2}
\beta_2:\begin{array}{lcl}
x_1&\to&x_3,\\
x_2&\to&-x_2-x_1x_2-\omega_2,\\
x_3&\to&x_1,\\
\end{array}\ee
\be\label{eq:braid3}
\beta_3:\begin{array}{lcl}
x_1&\to&x_2,\\
x_2&\to&x_1,\\
x_3&\to&-x_3-x_1 x_2 -\omega_3.\\
\end{array}\ee
Note that two of these are enough to generate the whole braid group.

We are now going to show that when $G_\infty=2$ (geometrically this means that we have a puncture at infinity), the action of the braid group coincides with a {\it tagged cluster algebra structure} \cite{ChS}.

In order to see this let us compose each braid with a Okamoto symmetry in order to obtain the following
\be\label{eq:braid-generic}
\widetilde \beta_i:\begin{array}{lcl}
x_i&\to&-x_i-x_j x_k -\omega_i,\quad j,k\neq i,\\
x_j&\to&x_j,\quad\hbox{for}\quad j\neq i\\
\end{array}\ee
By using (\ref{eq:mon-mf}) this transformation acquires a cluster flavour:
\be\label{eq:braid-cluster}
\widetilde \beta_i: \, x_ix_i'=x_j^2+x_k^2+\omega_j x_j+\omega_k x_k +\omega_4\quad j,k\neq i.\\
\ee
Indeed let us introduce the shifted variables:
$$
y_i:=x_i-G_i,\quad i=1,2,3,
$$
they satisfy the {\it tagged cluster algebra relation:}
\be\label{eq:tagcluster}
\mu_i: \, y_iy_i'=y_j^2+y_k^2+G_i y_j y_k\quad j,k\neq i.\\
\ee
Note that tagged cluster algebras satisfy the Laurent phenomenon. In particular this result implies that procedure of analytic continuation of the solutions to the sixth Painlev\'e equation satisfies the Laurent phenomenon: if we start from a local solution corresponding to the point $(y_1^0,y_2^0,y_3^0)$ on the shifted Painlev\'e cubic
$$
y_1 y_2 y_3 + y_1^2+y_2^2 + y_3^2 + G_1 y_2 y_3+G_2 y_1 y_3+G_3 y_1 y_2=0
$$
any other branch of that solution will corresponds to points   $(y_1,y_2,y_3)$ on the same cubic such that each $y_i$ is a Laurent polynomial of the initial  $(y_1^0,y_2^0,y_3^0)$.

\begin{remark}
A similar tagged cluster algebra structure can be found also for the fifth and the third Painlev\'e equation (see Subsections \ref{suse:tag1} and \ref{suse:tag1} below). However the meaning of this tagged cluster algebra structure in terms of analytic continuation of the solutions is still to be clarified and is postponed to subsequent publications.
\end{remark}

\subsection{Tagged cluster algebra structure for PV and PIII}\label{suse:tag1}

In this case, only $x_1$ and $x_2$ can be mutated. In the case of the $\widetilde D_5$ cubic, 
the formula for these mutation is the same as before:
\be\label{eq:braid-generic1}
\widetilde \beta_i:\begin{array}{lcl}
x_i&\to&-x_i-x_j x_k -\omega_i,\quad j,k\neq i,\\
x_j&\to&x_j,\quad\hbox{for}\quad j\neq i\\
\end{array}\quad i=1,2,\ee
where $\omega_i=\omega_i^{(\widetilde D_5)}$. 
The the shifted variables:
$$
y_i:=x_i+t_i,\quad i=1,2,3,
$$
where 
$$
t_1= -\frac{G_\infty(G_2 G_\infty-G_1)}{G_\infty^2-1},\quad t_2= -\frac{G_\infty(G_1 G_\infty-G_2)}{G_\infty^2-1},\quad 
t_3=-\frac{1+G_\infty^2}{G_\infty},
$$
satisfy the tagged cluster algebra:
\bea\label{eq:tagcluster5}
&&
\mu_1: \, y_1y_1'=y_2^2-t_1 y_2 y_3+ \nu y_3, \nn\\
&&
\mu_2: \, y_2y_2'=y_1^2-t_2 y_1 y_3+ \nu y_3, \nn
\eea
where
$$
\nu=-\frac{G_\infty(1+G_1^2 G_\infty^2-(2-G_2^2) G_\infty^2+G_\infty^4-G_1 G_2 G_\infty(1+ G_\infty^2)}{(G_\infty^2-1)^2}.
$$
Analogous computations can be repeated for PIII.

\subsection{Tagged cluster algebra structure for PIV}\label{suse:tag1}

In the case of the $\widetilde E_6$ cubic, only $x_1$ can be mutated. The formula for this mutation is the same as usual:
\be\label{eq:braid-generic4}
\widetilde \beta:\begin{array}{lcl}
x_1&\to&-x_1-x_2 x_3 -\omega_1,\\
x_j&\to&x_j,\quad\hbox{for}\quad j=2,3\\
\end{array}\ee
where $\omega_i=\omega_i^{(\widetilde E_6)}$. 
We present here the shifted variables for the case $G_\infty=1$:
$$
y_i:=x_i+t_i,\quad i=1,2,3,
$$
where 
$$
t_1= -1+\frac{5}{4 G_1},\quad t_2= \frac{1}{2}-\frac{5}{4 G_1},\quad 
t_3=-2,
$$
satisfy the tagged cluster algebra:
\be\label{eq:tagcluster3}
\mu_1: \, y_1y_1'=y_2^2-t_1 y_2 y_3+ \nu y_3,
\ee
where
$$
\nu=-\frac{25-30 G_1+ 24 G_1^2}{32 G_1^2}.
$$

\section{Shear coordinates for the Painlev\'e monodromy manifolds}\label{se:shear}

In the $D_4$ case the following parameterisation of the cubic in shear coordinates on the fat-graph of a $4$--holed sphere was found in \cite{ChM}:
\begin{eqnarray}
\label{eq:shear-PVI}
&&
x_1=-e^{\tilde s_2+ \tilde s_3}-e^{-\tilde  s_2-\tilde  s_3}-e^{- \tilde s_2+ \tilde s_3}-G_2e^{\tilde s_3}-G_3 e^{-\tilde s_2}\nn\\
&&
x_2=-e^{\tilde s_3+\tilde s_1}-e^{-\tilde s_3-\tilde s_1}-e^{-\tilde s_3+\tilde s_1}-G_3e^{\tilde s_1}-G_1e^{-\tilde s_3},\\
&&
x_3=-e^{\tilde s_1+ \tilde s_2}-e^{-\tilde s_1- \tilde s_2}-e^{-\tilde s_1+\tilde  s_2}-G_1e^{\tilde s_2}-G_2 e^{- \tilde s_1}\nn
\end{eqnarray}
where 
$$
G_i=e^{\frac{p_i}{2}}+e^{-\frac{p_i}{2}},\qquad i=1,2,3,
$$
and 
$$
G_\infty=e^{\tilde s_1+\tilde s_2+\tilde s_3}+e^{-\tilde s_1-\tilde s_2-\tilde s_3},
$$
and $\tilde s_i$ are actually the shifted shear coordinates $\tilde s_i=s_i +\frac{p_i}{2}$, $i=1,2,3$.

We recall that according to Fock~\cite{Fock1}~\cite{Fock2}, 
the fat graph associated to a Riemann
surface  $\Sigma_{g,n}$ of genus $g$ and with $n$ holes is a connected three--valent
graph drawn without self-intersections on $\Sigma_{g,n}$
with a prescribed cyclic ordering
of labelled edges entering each vertex; it must be a maximal graph in
the sense that its complement on the Riemann surface is a set of
disjoint polygons (faces), each polygon containing exactly one hole
(and becoming simply connected after gluing this hole).
In the case of a Riemann sphere $\Sigma_{0,4}$ with $4$ holes, the fat--graph is represented in Fig.1. 

The geodesic length functions, which are traces of hyperbolic elements in the Fuchsian group $\Delta_{g,s}$ such that 
$$
\Sigma_{g,s}\sim\mathbb H\slash \Delta_{g,s}
$$
are obtained by decomposing each hyperbolic matrix $\gamma\in \Delta_{g,s}$ into a
product of the so--called {\it right, left and edge matrices:}
\begin{equation}\nn
R:=\left(\begin{array}{cc}1&1\\-1&0\\
\end{array}\right), \qquad
X_{s_i}:=\left(\begin{array}{cc}0&-\exp\left({\frac{s_i}{2}}\right)\\
\exp\left(-{\frac{s_i}{2}}\right)&0\end{array}\right).
\label{eq:generators}
\end{equation}
In this setting our $x_1,x_2x_3$ are the geodesic lengths of thee geodesics which go around two holes without self--intersections, for example $x_3$ corresponds to the dashed geodesic in Fig.1.

\begin{figure}
\label{loopinvert}
{\psset{unit=0.5}
\begin{pspicture}(-5,-5)(7,5)
\newcommand{\PATTERN}[1]{%
\pcline[linewidth=1pt](0.3,0.5)(2,0.5)
\pcline[linewidth=1pt](0.3,-0.5)(2,-0.5)
\psbezier[linewidth=1pt](2,0.5)(3,2)(5,2)(5,0)
\psbezier[linewidth=1pt](2,-0.5)(3,-2)(5,-2)(5,0)
\psbezier[linewidth=1pt](2.8,0)(3.6,0.8)(4,0.7)(4,0)
\psbezier[linewidth=1pt](2.8,0)(3.6,-0.8)(4,-0.7)(4,0)
\rput(1,1.2){\makebox(0,0){$s_{#1}$}}
\rput(4.5,2){\makebox(0,0){$p_{#1}$}}
}
\rput(0,0){\PATTERN{1}}
\rput{120}(0,0){\PATTERN{2}}
\rput{240}(0,0){\PATTERN{3}}
\newcommand{\CURVE}{%
\pcline[linecolor=red, linestyle=dashed, linewidth=1.5pt](0.1,.2)(2.1,.2)
\pcline[linecolor=red, linestyle=dashed, linewidth=1.5pt](0.1,-.2)(2.1,-.2)
\psbezier[linecolor=red, linestyle=dashed, linewidth=1.5pt](2.1,.2)(3,1.7)(4.7,1.8)(4.7,0)
\psbezier[linecolor=red, linestyle=dashed, linewidth=1.5pt]{->}(2.1,-.2)(3,-1.6)(4.7,-1.8)(4.7,0)
}
\rput(0,0){\CURVE}
\rput{120}(0,0){\CURVE}
\psarc[linecolor=red, linestyle=dashed, linewidth=1.5pt](0.1,.2){.4}{210}{270}
\end{pspicture}
}\caption{The fat graph of the $4$ holed Riemann sphere. The dashed geodesic corresponds to $x_3$.}
\end{figure}
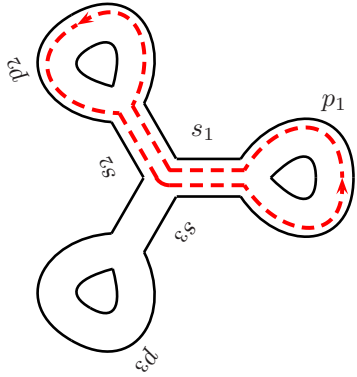

In \cite{ChM} it was shown that flips on the shear coordinates correspond to the action of the braid group on the cubic. The flips of the shear coordinates which give rise to the braid transformations $\widetilde\beta_1,\widetilde\beta_2$ and $\widetilde\beta_3$  have the following form
\be\label{eq:flip1}
f_1:\begin{array}{lcl}
\tilde s_1&\to&\tilde s_1,\\
\tilde s_2&\to&-\tilde s_2-\log\left[1+G_1 e^{\tilde s_1}+e^{2\tilde s_1}\right],\\
\tilde s_3&\to&-\tilde s_3+\log\left[1+G_1 e^{-\tilde s_1}+e^{-2\tilde s_1}\right],\\
\end{array}\ee
\be\label{eq:flip2}
f_2:\begin{array}{lcl}
\tilde s_1&\to&-\tilde s_1+\log\left[1+G_2 e^{-\tilde s_2}+e^{-2\tilde s_2}\right],\\
\tilde s_2&\to&\tilde s_2,\\
\tilde s_3&\to&-\tilde s_3-\log\left[1+G_2 e^{\tilde s_2}+e^{2\tilde s_2}\right],\\
\end{array}\ee
\be\label{eq:flip3}
f_3:\begin{array}{lcl}
\tilde s_1&\to&-\tilde s_1-\log\left[1+G_3 e^{\tilde s_3}+e^{2\tilde s_3}\right],\\
\tilde s_2&\to&-\tilde s_2+\log\left[1+G_3 e^{-\tilde s_3}+e^{-2\tilde s_3}\right],\\
\tilde s_3&\to&\tilde s_3.\\
\end{array}\ee

\begin{remark}
Observe that in \cite{ChS} it was proved that shear coordinate flips (\ref{eq:flip1}), (\ref{eq:flip2}),  (\ref{eq:flip3}) are indeed {\it dual}\/ to the tagged cluster mutations (\ref{eq:tagcluster}) for the corresponding $\lambda$-lengths. 
\end{remark}

We are now going to produce a similar shear coordinate description of each of the other Painlev\'e cubics. For $\widetilde D_5,\widetilde D_6,\widetilde E_6,\widetilde E_7$ we will provide a geometric description of the corresponding Riemann surface and its fat-graph. Our geometric description agrees with the one obtained in \cite{S}, which was obtained by building a Strebel differential  from the isomonodromic problems associated to each of the Painlev\'e equations.

\subsection{Shear coordinates for $\widetilde D_5$}

The confluence from the cubic associated to PVI to the one associated to PV is realised by 
$$
\tilde s_3\to \tilde s_3-\log[\epsilon],\qquad p_3\to p_3 -2\log[\epsilon],
$$
in the limit $\epsilon\to0$.  We obtain the following shear coordinate description for the $\widetilde D_5$ cubic:
\begin{eqnarray}
\label{eq:shear-PV}
&&
x_1=-e^{\tilde  s_2+\tilde  s_3}-e^{- \tilde s_2+\tilde  s_3}-G_2e^{\tilde s_3}-G_3 e^{-\tilde s_2}\nn\\
&&
x_2=-e^{\tilde s_3+\tilde s_1}-G_3e^{\tilde s_1},\\
&&
x_3=-e^{\tilde s_1+ \tilde s_2}-e^{-\tilde s_1-\tilde s_2}-e^{-\tilde s_1+ \tilde s_2}-G_1e^{\tilde s_2}-G_2 e^{-\tilde s_1}\nn
\end{eqnarray}
where 
$$
G_i=e^{\frac{p_i}{2}}+e^{-\frac{p_i}{2}},\quad i=1,2,
\quad G_3=e^{\frac{p_3}{2}},\quad
G_\infty=e^{\tilde s_1+\tilde s_2+\tilde s_3}.
$$
To obtain the cubic in our form we need to specialise $p_3=0$.

To understand the geometry of this confluence we need to revert to the non--shifted shear coordinates, in which the confluence is realised by 
$$
s_3\to s_3, \qquad p_3\to p_3 -2\log[\epsilon].
$$
This means that we send the perimeter $p_3$ to infinity, which is the same as opening one of the faces in two infinite directions as in Fig.2. Geometrically speaking this correspond to a Riemann sphere with three holes and two marked points on one of them.

\begin{figure}
\label{loopinvert}
{\psset{unit=0.5}
\begin{pspicture}(-5,-5)(7,5)
\newcommand{\PATTERN}[1]{%
\pcline[linewidth=1pt](0.3,0.5)(2,0.5)
\pcline[linewidth=1pt](0.3,-0.5)(2,-0.5)
\psbezier[linewidth=1pt](2,0.5)(3,2)(5,2)(5,0)
\psbezier[linewidth=1pt](2,-0.5)(3,-2)(5,-2)(5,0)
\psbezier[linewidth=1pt](2.8,0)(3.6,0.8)(4,0.7)(4,0)
\psbezier[linewidth=1pt](2.8,0)(3.6,-0.8)(4,-0.7)(4,0)
\rput(1,1.2){\makebox(0,0){$s_{#1}$}}
\rput(4.5,2){\makebox(0,0){$p_{#1}$}}
}
\rput(0,0){\PATTERN{1}}
\rput{120}(0,0){\PATTERN{2}}
\rput{240}(0,0){\PATTERN{3}}
\pscircle[linecolor=white,fillstyle=solid](-3,-3.5){1.5}
\pscircle[linecolor=white,fillstyle=solid](-1,-4.2){1.3}
\newcommand{\CURVE}{%
\pcline[linecolor=red, linestyle=dashed, linewidth=1.5pt](0.1,.2)(2.1,.2)
\pcline[linecolor=red, linestyle=dashed, linewidth=1.5pt](0.1,-.2)(2.1,-.2)
\psbezier[linecolor=red, linestyle=dashed, linewidth=1.5pt](2.1,.2)(3,1.7)(4.7,1.8)(4.7,0)
\psbezier[linecolor=red, linestyle=dashed, linewidth=1.5pt]{->}(2.1,-.2)(3,-1.6)(4.7,-1.8)(4.7,0)
}
\end{pspicture}
}\caption{The fat graph of a $3$-holed Riemann sphere with two marked points on one boundary component.}
\end{figure}
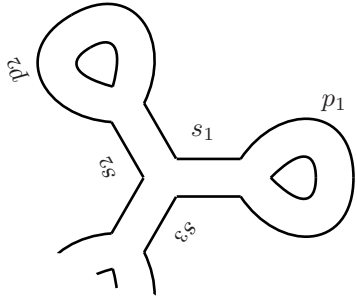

\subsection{Shear coordinates for $\widetilde E_6$}

The confluence from PV to PIV is realised by the substitution
$$
\tilde s_2\to\tilde s_2-\log[\epsilon],\qquad p_2\to p_2 -2\log[\epsilon],
$$
in formulae (\ref{eq:shear-PV}). In the limit $\epsilon\to0$ we obtain:
\begin{eqnarray}
\label{eq:shear-PIV}
&&
x_1=-e^{ \tilde s_2+ \tilde s_3}-G_2e^{\tilde s_3}\nn\\
&&
x_2=-e^{\tilde s_3+\tilde s_1}-G_3e^{\tilde s_1},\\
&&
x_3=-e^{\tilde s_1+ \tilde s_2}-e^{-\tilde s_1+ \tilde s_2}-G_1e^{\tilde s_2}-G_2 e^{- \tilde s_1}\nn
\end{eqnarray}
where 
$$
G_1=e^{\frac{p_1}{2}}+e^{-\frac{p_1}{2}},\quad 
G_i=e^{\frac{p_i}{2}},\, i=2,3,\quad
G_\infty=e^{\tilde s_1+\tilde s_2+\tilde s_3}.
$$
To obtain the cubic in our form we need to specialise $p_3=p_2=2\tilde s_1+2\tilde s_2+2\tilde s_3$.

Again, to understand the geometry of this confluence we need to revert to the non--shifted shear coordinates, in which the confluence is realised by 
$$
s_2\to s_2, \qquad p_2\to p_2 -2\log[\epsilon].
$$
Similarly to the previous case, this means that we send the perimeter $p_2$ to infinity, which is the same as opening one of the two faces without marked points in two infinite directions. Geometrically speaking this correspond to a Riemann sphere with two holes, one of which has $4$ marked points.

\subsection{Shear coordinates for $\widetilde E_7$}

The confluence from PIV to PII is realised by the substitution
$$
\tilde s_1\to \tilde s_1-\log[\epsilon],\qquad p_1\to p_1 -2\log[\epsilon],
$$
in formulae (\ref{eq:shear-PIV}). In the limit $\epsilon\to0$ we obtain:
\begin{eqnarray}
\label{eq:shear-PII}
&&
x_1=-e^{\tilde s_2+\tilde s_3}-G_2e^{\tilde s_3}\nn\\
&&
x_2=-e^{\tilde s_3+\tilde s_1}-G_3e^{\tilde s_1},\\
&&
x_3=-e^{\tilde s_1+ \tilde s_2}-G_1e^{\tilde s_2}\nn
\end{eqnarray}
where 
$$
G_i=e^{\frac{p_i}{2}},\, i=1,2,3,\quad
G_\infty=e^{\tilde s_1+\tilde s_2+\tilde s_3}.
$$
To obtain the $\widetilde E_7^\ast$ cubic  we need to specialise $p_3=p_2=p_1=-2(\tilde s_1+\tilde s_2+\tilde s_3)$, while to get the $\widetilde E_7^{\ast\ast}$ one we need $p_3=-p_2=p_1=-2(\tilde s_1+\tilde s_2+\tilde s_3)$.

Again geometrically this confluence gives rise to a Riemann sphere with one hole and $6$ marked points.

\subsection{Shear coordinates for $\widetilde E_7$}
The confluence from PII to PI is realised by 
$$
\tilde s_3\to \tilde s_3-\log[\epsilon],\qquad p_3\to p_3 +2\log[\epsilon],
$$
in formulae (\ref{eq:shear-PII}). In t the limit $\epsilon\to0$ we obtain:
\begin{eqnarray}
\label{eq:shear-PI}
&&
x_1=-e^{ \tilde s_2+ \tilde s_3}-G_2e^{\tilde s_3}\nn\\
&&
x_2=-e^{\tilde s_3+\tilde s_1},\\
&&
x_3=-e^{\tilde s_1+ \tilde s_2}-G_1e^{\tilde s_2}\nn
\end{eqnarray}
where 
$$
G_i=e^{\frac{p_i}{2}},\, i=1,2,\quad G_3=0,\quad
G_\infty=e^{\tilde s_1+\tilde s_2+\tilde s_3}.
$$
To obtain the cubic in our form we need to specialise $p_2=p_1=0$ and $\tilde s_1+\tilde s_2+\tilde s_3=0$.

\subsection{Shear coordinates for $\widetilde D_6$}
The confluence from PV to PIII $\widetilde D_6$ is realised by the substitution
$$
\tilde s_3\to \tilde s_3-\log[\epsilon],\qquad p_3\to p_3+2\log[\epsilon],
$$
in formulae (\ref{eq:shear-PV}). in the limit $\epsilon\to0$ we obtain:
\begin{eqnarray}
\label{eq:shear-PIII}
&&
x_1=-e^{\tilde s_2+ \tilde s_3}-e^{- \tilde s_2+ \tilde s_3}-G_2e^{\tilde s_3}\nn\\
&&
x_2=-e^{\tilde s_3+\tilde s_1},\\
&&
x_3=-e^{\tilde s_1+ \tilde s_2}-e^{-\tilde s_1+ \tilde s_2}-e^{-\tilde s_1-\tilde s_2}-G_1e^{\tilde s_2}-G_2 e^{- \tilde s_1}\nn
\end{eqnarray}
where 
$$
G_i=e^{\frac{p_i}{2}}+e^{-\frac{p_i}{2}},\quad i=1,2,\quad G_3=0,\quad
G_\infty=e^{\tilde s_1+\tilde s_2+\tilde s_3}.
$$
To obtain the cubic in our form we need to specialise $p_1=2\tilde s_1+2\tilde s_2+2\tilde s_3$. 

To understand this confluence from a geometric point of view we first revert to the unshifted shear coordinates, which are transformed as follows:
$$
s_3\to s_3-\log[\epsilon^2],\qquad p_3\to p_3+2\log[\epsilon],
$$
the second part of this scaling corresponds to dragging back from $\infty$ the two infinite directions produced by the PVI to PV confluence, and clashing them into a point with $G_3=0$. The first part of the scaling sends this point to infinity. This corresponds to a Riemann sphere with three holes and one marked point on one of them. The best way to understand this is to consider a Riemann sphere with $5$ holes and an involution which identifies two couples of opposite holes and rotates the fifth (see figure 3). This involution admits an   invariant curve with a stable point on it. We select a geodesic homothopic to this curve and cut along it, obtaining then a Riemann sphere with three holes and a marked point (the projection of the stable one).\footnote{The authors are grateful to L. Chekhov for his insights on the geometric interpretation of this confluence.}

\begin{figure}
\begin{center}
\begin{pspicture}(-3,-3)(3,3)
 \psellipse[linecolor=red](-2,1.5)(0.5,0.2)
  \psellipse[linecolor=blue](2,1.5)(0.5,0.2)
   \psellipse[linecolor=red](-2,-1.5)(0.5,0.2)
  \psellipse[linecolor=blue](2,-1.5)(0.5,0.2)
    \psellipse(3,0)(0.2,0.5)
\psarc(-8.3,0){6}{-14}{14}
\psarc(0,7.4){6}{254}{287}
\psarc(0,-7.35){6}{75}{105}
\psarc(4.5,2){2.1}{194}{228}
\psarc(4.5,-2){2.1}{133}{168}
\psarc[linecolor=green](0,10.5){10}{256.5}{286}
\psline[linestyle=dashed](-2.3,0)(2.8,0)
\rput(-2.3,0){$\bullet$}
\end{pspicture}
\caption{Geodesics of the same color are identified. The dashed line corresponds to the invariant curve and its stable point. The green gedesic is the one homothopic to it.}
\end{center}
\end{figure}
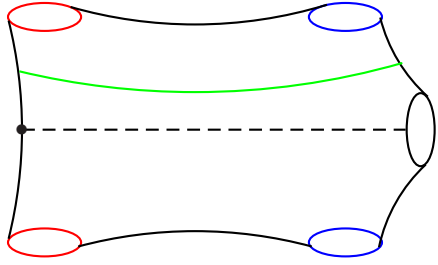

\subsection{Shear coordinates for $\widetilde D_7$}
The confluence from PV to PIII $\widetilde D_7$ is realised by the substitution
\bea
\tilde s_1\to \tilde s_1-\log[\epsilon],& \tilde s_2\to \tilde s_2+\log[\epsilon],& p_2\to p_2 -2\log[\epsilon],\nn\\
\tilde s_3\to \tilde s_3-\log[\epsilon],& p_3\to p_3-2\log[\epsilon],&\nn
\eea
in formulae (\ref{eq:shear-PV}). in the limit $\epsilon\to0$ we obtain:
\begin{eqnarray}
\label{eq:shear-PIIID7}
&&
x_1=-e^{-\tilde  s_2+ \tilde s_3}-G_2e^{\tilde s_3}-G_3 e^{-\tilde s_2}\nn\\
&&
x_2=-e^{\tilde s_3+\tilde s_1}-G_3e^{\tilde s_1},\\
&&
x_3=-e^{\tilde s_1+\tilde  s_2}-e^{-\tilde s_1-\tilde s_2}-G_2 e^{- \tilde s_1}\nn
\end{eqnarray}
where 
$$
G_i=e^{\frac{p_i}{2}},\quad i=1,2,3,\quad
G_\infty=e^{\tilde s_1+\tilde s_2+\tilde s_3}.
$$
To obtain the cubic in our form we need to specialise $p_1=-2(\tilde s_1+\tilde s_2+\tilde s_3)$.

\subsection{Shear coordinates for $\widetilde D_8$}
The confluence from PV to PIII $\widetilde D_8$ is realised by the substitution
$$
s_1\to s_1-\log[\epsilon], s_2\to s_2+\log[\epsilon], p_2\to p_2 -2\log[\epsilon],
$$
in formulae (\ref{eq:shear-PIII}). In the limit $\epsilon\to0$ we obtain:
\begin{eqnarray}
\label{eq:shear-PIIID8}
&&
x_1=-e^{- \tilde s_2+ \tilde s_3}-G_2e^{\tilde s_3}\nn\\
&&
x_2=-e^{\tilde s_3+\tilde s_1},\\
&&
x_3=-e^{\tilde s_1+ \tilde s_2}-e^{-\tilde s_1-\tilde s_2}-G_2 e^{- \tilde s_1}\nn
\end{eqnarray}
where 
$$
G_1=G_3=0,\quad G_2=e^{\frac{p_2}{2}},\quad
G_\infty=e^{\tilde s_1+\tilde s_2+\tilde s_3}.
$$
To obtain the cubic in our form we need to specialise $G_2=0$.

\section{Quantisation}\label{se:q}

In this section we provide the quantisation of all the Painlev\'e cubics and produce the corresponding {\it quantum confluence} in such a way that quantisation and confluence commute.

As discussed in Section \ref{se:volume-forms}, on each Painlev\'e cubic surface denoted by an index $d$ running on the list of the extended Dynkin diagrams $\widetilde D_4,\widetilde D_5,\widetilde D_6, \widetilde D_7,\widetilde D_8,\widetilde E_6,\widetilde E_7^\ast,\widetilde E_7^{\ast\ast}$, $\widetilde E_8$, we have the following Poisson bracket:
\be\label{eq:poisson}
\{x_i,x_{i+1}\}= x_i x_{i+1}+ 2\epsilon^{(d)}_{k} x_k+\omega^{(d)}_k,\quad k\neq i,i+1,
\ee
where we use the cyclic notation $x_{i+3}=x_i$, $i=1,2,3$ and the parameters $\epsilon^{(d)}_{i,2,3}$, and $\omega_{1,2,3}^{(d)}$ are given by the formulae (\ref{eq:epsilon}) and (\ref{eq:omega}) respectively.

This Poisson algebra is induced by the Poisson algebras of geodesic length functions constructed in
\cite{Ch1} by  postulating the Poisson relations on the level of the
shear coordinates $s_\alpha$ of the Teichm\"uller space. In our case these are:
$$
\{s_1,s_2\}=\{s_2,s_3\}=\{s_3,s_1\}=1,
$$ 
while the perimeters $p_1,p_2,p_3$ are assumed to be Casimirs so that the shifted shear coordinates $\tilde s_1,\tilde s_2, \tilde s_3$ satisfy the same Poisson relations:
$$
\{\tilde s_1,\tilde s_2\}=\{\tilde s_2,\tilde s_3\}=\{\tilde s_3,\tilde s_1\}=1.
$$
It is worth reminding that the exponentials of the shear coordinates satisfy the log-canonical Poisson bracket. 

To produce the quantum Painlev\'e cubics, we introduce the Hermitian operators $S_1,S_2,S_3$ subject to the commutation
inherited from the Poisson bracket of $\tilde s_i$:
$$
[S_i,S_{i+1}]=i\pi \hbar \{\tilde s_i,\tilde s_{i+1}\}=i\pi \hbar,\quad i=1,2,3,\ i+3\equiv i.
$$
Observe that thanks to this fact, the commutators $[S_i,S_{j}]$ are always numbers and therefore we have
$$
\exp\left({a S_i}\right) \exp\left({b S_j}\right) =
\exp\left(a {S_i}+b {S_i}+\frac{ab}{2}[S_i,S_{j}]\right) ,
$$
for any two constants $a,b$. Therefore we have the Weyl ordering:
$$
e^{S_{1}+S_{2}}=q^{\frac{1}{2}}e^{S_{1}}e^{S_{2}}=q^{-\frac{1}{2}}e^{S_{2}}e^{S_{1}},\quad q\equiv e^{-i\pi\hbar}.
$$
After quantisation, the perimeters $p_1,p_2,p_3$  and $\tilde s_1+\tilde s_2+\tilde s_3$ remain non--deformed, so 
we preserve the previous notation for
them. This is equivalent to say that the constants $\omega_i^{(d)}$ remain non-deformed. 

We introduce the Hermitian operators $X_1,X_2,X_3$ as follows: consider the classical expressions for $x_1,x_2,x_3$ is terms of $\tilde s_1,\tilde s_2,\tilde s_3$ and $p_1,p_2,p_3$. Write each product of exponential terms as the exponential of the sum of the exponents and replace those exponents by their quantum version. For example, in the case of $\widetilde D_5$ we have:
$$
x_1=-e^{\tilde  s_2+\tilde  s_3}-e^{- \tilde s_2+\tilde  s_3}-G_2e^{\tilde s_3}-G_3 e^{-\tilde s_2},
$$
and its quantum version is defined as
\bea
X_1&=&-e^{-S_2}-(e^{\frac{p_2}{2}}+e^{-\frac{p_2}{2}})e^{S_3}-e^{-S_2+S_3}-e^{S_2+S_3}=\nn\\
&=&
e^{-S_2}-(e^{\frac{p_2}{2}}+e^{-\frac{p_2}{2}})e^{S_3}-q^{-\frac{1}{2}}e^{-S_2}e^{S_3}-q^{\frac{1}{2}}e^{S_2}e^{S_3}.\nn
\eea
\begin{theorem} Denote by $X_1,X_2,X_3$ the quantum Hermitian operators corresponding to $x_1,x_2,x_3$ as above. The quantum commutation relations are:
\be\label{eq:q-comm}
q^{\frac{1}{2}} X_i X_{i+1}- q^{-\frac{1}{2}} X_{i+1} X_i = \left(\frac{1}{q}-q\right) \epsilon^{(d)}_{k}  X_k+
(q^{-\frac{1}{2}}-q^{\frac{1}{2}}) \omega_k^{(d)}
\ee
where $\epsilon^{(d)}_i$ and $\omega_i^{(d)}$ are the same as in the classical case. The quantum operators satisfy the following quantum cubic relations:
\be\label{q-cubics}
q^{\frac{1}{2}} X_3 X_1 X_2+ q X^2_3 + q^{-1}  \epsilon^{(d)}_1 X_1^2+ q  \epsilon^{(d)}_2 X_2^2 +
 q^{-\frac{1}{2}}  \epsilon^{(d)}_3+ \omega_3 X_3 + q^{\frac{1}{2}}  \omega^{(d)}_1 X_1 + q^{\frac{1}{2}}  \omega^{(d)}_2 X_2+ \omega_4^{(d)}=0.
\ee
\end{theorem}

\begin{remark}Observe that in the case of PVI the above quantum commutation relations (\ref{eq:q-comm}) and quantum cubic (\ref{q-cubics})  appeared already in the paper by Ito and Terwillinger \cite{IT} (see also \cite{EE}) in relation with the spherical sub-algebra of  the generalised rank $1$ double affine Hecke (DAHA) algebra studied in
\cite{Sa}. Their work followed a theorem by  Oblomkov
\cite{Obl} in which the  classical cubic (\ref{eq:mon-mf}) for the $\widetilde D_4$ case appeared as the spectrum of the centre 
of the same generalised DAHA.\end{remark}

\begin{remark} In the case of PII the quantum commutation relations (\ref{eq:q-comm}) (up to re-scaling) coinside with the equitable presentation of $U_q(sl_2)$, due to Ito, Terwilliger and Weng \cite{ITW}. This algebra is generated by $x, y $ and $z^{\pm1}$ subject to the relations
$$q^2xy - yx = q^2 -1, q^2yz - zy = q^2 - 1\quad  \text{and}\quad  q^2zx - xz = q^2 - 1.$$\end{remark}

\begin{remark}It is clear that if we define the quantum confluence as the obvious analogue of the classical one, i.e. we rescale the quantum Hermitian operators by a constant $\epsilon$ and take the limit as $\epsilon\to 0$, then quantisation and confluence commute.\end{remark}


\end{document}